\documentclass[final,5p,times,twocolumn]{elsarticle}
\usepackage{lineno,hyperref}
\usepackage[dvipsnames]{xcolor}
\usepackage{amsfonts}
\usepackage{comment}
\usepackage{graphicx}
\usepackage{subcaption}
\usepackage{amsmath, amssymb}
\newcommand{\for}{\text{for }}

\journal{Journal of \LaTeX\ Templates}

\begin{document}

\begin{frontmatter}

\title{FogBus: A Blockchain-based Lightweight Framework for Edge and Fog Computing}

\author{Shreshth Tuli\fnref{one,two}}
\fntext[one]{Cloud Computing and Distributed Systems (CLOUDS) Laboratory\\School of Computing and Information Systems\\The University of Melbourne, Australia\\Email: mahmudm@student.unimelb.edu.au}
\fntext[two]{Department of Computer Science,\\Indian Institute of Technology (IIT), Delhi, India}
\author{Redowan Mahmud \fnref{one}}
\author{Shikhar Tuli\fnref{three}}
\fntext[three]{Department of Electrical Engineering,\\Indian Institute of Technology (IIT), Delhi, India}
\author{Rajkumar Buyya\fnref{one}}

\begin{abstract}
The requirement of supporting both latency sensitive and computing intensive Internet of Things (IoT) applications is consistently boosting the necessity for integrating Edge, Fog and Cloud infrastructure. Although there are a number of real-world frameworks attempt to support such integration, they have many limitations from various perspectives including platform independence, security, resource management and multi-application assistance. To address these limitations, we propose a simplified but effective framework, named \textit{FogBus} for facilitating end-to-end IoT-Fog(Edge)-Cloud integration. FogBus offers a platform independent interface to IoT applications and computing instances for execution and interaction. It not only assists developers in building applications but also helps users in running multiple applications at a time and service providers to manage their resources. In addition, FogBus applies Blockchain, authentication and encryption techniques to secure operations on sensitive data. Because of its lightweight and cross platform software systems, it is easy to deploy, scalable and cost efficient. We demonstrate the effectiveness of our framework by creating a computing environment with it that integrates finger pulse oximeter as IoT devices with Smartphone-based gateway and Raspberry Pi-based Fog nodes for Sleep Apnea analysis. We also run several experiments on this computing environment varying FogBus settings. The experimental results show that different FogBus settings can improve latency, energy, network and CPU usage of the computing infrastructure.  
\end{abstract}


\end{frontmatter}


\section{Introduction}
Internet of Things (IoT) paradigm enables different sensors, digital machines and objects to perceive the external environment and connects them to the global Internet for exchanging data. It also supports integration and analysis of generated data through application software so that events of interest can be identified and necessary physical actions can be triggered through actuators. Thus, it paves the way for building smart systems with limited human intervention \cite{IoT}. Based on the current trends, it is expected that by 2025 such systems will incorporate over 1 trillion IoT devices with 50\% increased demand of latency sensitive applications \cite{mcKinsey}. So far, Cloud is considered as the fundamental computing paradigm to deal with these large number of geographically distributed IoT devices and host corresponding IoT applications. However, Cloud datacenters reside at multi-hop distance from IoT devices that increase communication delay in both transferring the data and receiving the application service \cite{speed}. For latency-sensitive applications such as healthcare and smart city, this high-latency interaction between IoT devices and Cloud datacenters is unacceptable and can degrade the service quality drastically. In addition, IoT devices can generate a huge amount of data within a minimal time. When a large number of IoT devices simultaneously initiate transferring the data to Cloud datacenters through global Internet, severe network congestion occurs. To overcome these limitations of Cloud-centric IoT model, Fog and Edge computing paradigms are emerged \cite{FogInit}. Both of them prefer to utilize edge resources provided by local computing instances for executing real-time IoT applications. Smart devices with computing processors such as Raspberry Pi devices, personal computers, mobile phones, network switches, routers and micro-datacenters can offer potential edge resources \cite{latencyAware}. Based on the capabilities of these edge resources, many consider Fog and Edge computing similar and use them interchangeably. Conversely, others treat Edge computing as a subset of Fog computing \cite{me}.                 
\par Fog computing manages an intermediate layer between IoT-enabled systems and Cloud computing. The computing instances of Fog computing are commonly known as Fog nodes and deployed across the edge network in distributed manner. Through these nodes, Fog provides Cloud-like services such as infrastructure, platform and software closer to the IoT data sources and supports application execution. Consequently, it reduces service delivery time and network congestion, and improves Quality of Service (QoS) and user experience \cite{QoE}. However, unlike Cloud datacenters, Fog nodes are resource constrained and heterogeneous. With limited resources, it is not possible to accommodate every compute intensive IoT application at the Fog layer. Therefore, seamless integration of IoT-enabled systems with Fog and Cloud infrastructure is required so that both edge and remote resources can be harnessed according to dynamics and requirements of the application \cite{interoperability}. In this integration, Cloud-centric top-down approach for managing Fog-based resources becomes infeasible when IoT-data is being received at a higher frequency for processing. On such occurrence, rather than relying on centralized resource management policies, it is effective to take decisions locally and provision resources following distributed bottom-up approach. Moreover, while placing and executing applications on this environment, both internal and external operations get obstructed by the heterogeneity of computing instances. In such circumstances, generalized techniques those are able to cope with varying contexts, can overcome the impediments of node to node communication and application runtime environment. Nevertheless, the implementation of an integrated environment going beyond the infrastructure and platform-level diversity with decentralized resource management policies is a challenging task. Its complexity is further intensified due to coexistence of multiple decision making entities, multi-dimensional scaling, unaware network topology and security issues \cite{ifogsim-Gupta}.   
\par In literature, there exists a number of works implementing software frameworks for integrating IoT-enabled systems, Fog and Cloud infrastructure \cite{amir} \cite{dubey} \cite{Yangui} \cite{Bruneo}. However, these frameworks barely support simultaneous execution of multiple applications and platform independence. Moreover, they offer narrow scope to application developers and users to tune the framework according to individual requirements. These frameworks exploit Cloud resources for data storage and often compel energy constrained IoT devices to process the raw data. To reduce the management overhead, existing frameworks apply centralized techniques that eventually degrade the QoS. They also confine the concentration on few security aspects which in consequence increases vulnerability of the integrated environment. In order to overcome such limitations of available frameworks, we develop a lightweight framework named \textit{FogBus}. 

\par FogBus allows an end-to-end implementation of integrated IoT-Fog-Cloud environment through a wide range of devices. FogBus provides platform independent application execution environment and node-to-node interaction. It assists developers in building applications, users in customizing services and providers in managing resources. Furthermore, FogBus facilitates execution of latency sensitive and compute intensive applications through Fog nodes and Cloud datacenters. Thus, it supports various types of applications simultaneously. To ensure data integrity, protection and privacy, FogBus also implements Blockchain and applies authentication and encryption techniques which consequently increase its reliability.\\
The \textbf{major contributions} of this work are listed as:
\begin{itemize}
\item Propose a lightweight framework named FogBus for integrating IoT enabled systems, Fog and Cloud infrastructure to harnesses both edge and remote resources according to application requirements.
\item Design of platform independent application execution and node-to-node interaction overcoming heterogeneity within the integrated environment.
\item Present a Platform-as-a-Service (PaaS) model that assists application developers, users and services providers to pursue individual interests. 
\item Development of a prototype application system for Sleep Apnea analysis in integrated IoT-Fog-Cloud environment.  
\item Implementation of Blockchain technique to ensure data integrity while transferring confidential data. 
\item Performance evaluation of FogBus in terms of latency, energy, network and CPU usage.            
\end{itemize} 
The rest of the paper is organized as follows. Section \ref{works} highlights key elements of several existing frameworks and compare them with our proposed framework. In Section \ref{mainPart}, the description of FogBus framework is provided. The design and implementation of FogBus are discussed in Section \ref{implementation}. In Section \ref{caseStudy} and \ref{performance}, a case study on Sleep Apnea analysis and performance of FogBus are presented respectively. Section \ref{future} concludes the paper proposing future works to improve FogBus.           
\section{Related Work} \label{works}
\begin{table*}[t]
\scriptsize
\centering 
\caption{Summary of the literature study}\label{Tab:summary} 
\begin{tabular}{|p{2.1 cm}|p{0.3cm}|p{0.3cm}|p{0.48cm}|p{1.8cm}|p{0.75 cm}|p{1.48cm}|p{1.1cm}|p{1.75cm}|p{1.1cm}|p{0.48cm}|p{1.50cm}|}
 \hline
 Work & \multicolumn{3} {c|}{Integrates} & Platform independent & \multicolumn{3} {c|}{Security features} & Supports multi-applications & \multicolumn{2} {c|}{Targets} & Decentralized management \\
\cline{2-4}
\cline{6-8}
\cline{10-11}
& IoT & Fog & Cloud & & Integrity & Authentication & Encryption & & Developers & Users &\\\hline
Amir et al. \cite{amir} & \checkmark & \checkmark & \checkmark & & & \checkmark & \checkmark & & & \checkmark & \checkmark \\\hline			Dubey et al. \cite{dubey} & \checkmark & \checkmark & & & & \checkmark & & \checkmark &  & \checkmark &  \checkmark \\\hline 
Azimi et al. \cite{azimi} & \checkmark & \checkmark & \checkmark & & & \checkmark & \checkmark & \checkmark &  & \checkmark &  \checkmark \\\hline 
Gia et al. \cite{Gia} & \checkmark & \checkmark & & & & \checkmark & \checkmark &  &  & \checkmark & \checkmark \\\hline 
Orestis et al. \cite{Orestis} & \checkmark & \checkmark & \checkmark & & & \checkmark & \checkmark &  &  & \checkmark & \\\hline  
Chen et al. \cite{chen} & \checkmark & \checkmark &  & & &  &  &  &  & \checkmark &  \checkmark \\\hline 
Razvan et al. \cite{razvan} & \checkmark & \checkmark &  & & &  &  &  &  & \checkmark &  \checkmark \\\hline  
Hu et al. \cite{Hu} & \checkmark & \checkmark & \checkmark & & & & &  &  & \checkmark &  \\\hline
Yangui et al. \cite{Yangui} & \checkmark & \checkmark & \checkmark & \checkmark & & \checkmark & \checkmark & \checkmark & \checkmark  &  & \\\hline  
Bruneo et al. \cite{Bruneo} & \checkmark & \checkmark &  & \checkmark & & \checkmark & \checkmark & \checkmark & \checkmark  &  & \\\hline  
Verba et al. \cite{verba} & \checkmark & \checkmark &  & \checkmark & & \checkmark & \checkmark & \checkmark & \checkmark  &  & \checkmark \\\hline
Yi et al. \cite{Yi} & \checkmark & \checkmark & \checkmark & \checkmark & & \checkmark &  & \checkmark & \checkmark  &  & \checkmark \\\hline 
Korosh et al. \cite{Korosh} & \checkmark & \checkmark &  & \checkmark & &  & \checkmark & \checkmark & \checkmark  & \checkmark & \checkmark \\\hline  
Chang et al. \cite{Chang} & \checkmark & \checkmark & \checkmark & \checkmark & \checkmark & \checkmark &  & \checkmark & \checkmark  &  & \\\hline 
Nader et al. \cite{nader} & \checkmark & \checkmark & \checkmark & \checkmark & \checkmark & \checkmark &  & \checkmark & \checkmark  &  & \checkmark \\\hline 
FogBus [this work] & \checkmark & \checkmark & \checkmark & \checkmark & \checkmark & \checkmark & \checkmark  & \checkmark & \checkmark  &  \checkmark & \checkmark \\\hline 
\end{tabular}  
\end{table*}

The existing frameworks that integrate different IoT-enabled systems with Fog and Cloud infrastructure are roughly classified into two types. The first type focuses on application specific prototypes while the other offers generalized PaaS model. Table \ref{Tab:summary} provides a brief summary of these frameworks.       
\par Amir et al. \cite{amir} develop a prototype-based framework for IoT-enabled health-care system,  enlightening the architecture of a smart gateway for facilitating local storage and data processing. In this framework, Cloud acts as a backend system for data analysis and decision making. To strengthen the framework, security featues of operating systems are used. Another prototype framework for smart health-care is developed by Dubey et al. \cite{dubey}. Intel Edison and Raspberry Pi circuits are used in the framework as Fog nodes. Through role-based authentication, the framework ensures privacy of the data. In this framework, Cloud is partially adopted for storing the data. Azimi et al. \cite{azimi} also discuss a hierarchical prototype framework for health-care solutions. The health analytics are divided into two parts to be placed separately in Cloud and Fog infrastructure. The framework follows MAPE-K model proposed by IBM to conduct the computations that inherently supports execution of diverse applications and provides encryption-based security.
\par Moreover, Gia et al. \cite{Gia} present a low-cost remote health monitoring framework that facilitates autonomic analysis of IoT data and notification generation. The IoT devices are designed with computational capabilities so that they can pre-process raw data and forward to the Fog nodes for further processing consuming less energy. In Fog layer distributed database is managed for data categorization and security. Orestis et al. \cite{Orestis} also develop a prototype framework that allows users to share health data and notify during emergency. Each operation within the framework is managed by a Spark IoT Platform Core residing at the Cloud. The framework uses encryption and authentication techniques for security. 
\par Chen et al. \cite{chen} and Razvan et al. \cite{razvan} develop separate prototype framework for smart city surveillance and gas-leak monitoring system respectively. In both frameworks the Fog infrastructure conducts necessary data processing and decision making operations. Remote Cloud is partially adopted in these frameworks for managing a record of IoT-data. However, the authors neither mention any security features nor techniques for managing heterogeneity of the Fog nodes within the framework. Likewise, Hu et al. \cite{Hu} omit the security issues of their developed prototype framework for face identification system. In this framework, a centralized Cloud manages all resources of integrated environment and offloads partial computational tasks to Fog infrastructure. After completing tasks on Fog, only the results are sent back to Cloud for further analysis and storage.
\par To promote IoT, Fog and Cloud integration, a Cloud-centric PaaS framework is developed by Yangui et al. \cite{Yangui} that automates the provisioning of applications. The PaaS facilitates developing diverse applications, their deployment and management of Fog nodes. The framework can deal with heterogeneity of the nodes. In addition, security features from Cloud Foundry architecture are extended in the framework. Similarly, Bruneo et al. \cite{Bruneo} propose a Fog-centric PaaS framework for deploying and executing multiple applications over computationally sound IoT devices. There, Fog infrastructure acts as a centralized programmable coordinator. The framework applies Cloud-based security features and deal with diverse applications surpassing heterogeneity of the instances. 
\par In addition Verba et al. \cite{verba} propose a gateway architecture that offers PaaS for integrating Fog nodes and IoT devices. The gateways assist messaging communication with authentication techniques. The framework supports horizontal integration of gateways and Cloud datacenters for application deployment and task migration. In another work Yi et al. \cite{Yi} propose a comprehensive PaaS framework for integrated environment. To implement the framework, resource-enriched Fog nodes, such as Cloudlet and ParaDrop, are required where each node including IoT devices should be virtualized. For securing the framework operations, existing authentication techniques are applied. Korosh et al. \cite{Korosh} also develop a PaaS framework that manages electricity usage in a home and in micro grid levels over Fog infrastructure. It can deal with the heterogeneity of Fog nodes and IoT devices and ensure data encryption.
\par The PaaS framework proposed by Chang et al. \cite{Chang} utilizes user's networking devices to execute IoT applications. In this framework, core services and resource management instructions are extended from Cloud datacenters to Fog nodes based on application requirements. It supports Cluster of Fog nodes and incorporates user's hand-held devices as well. For security, it runs a registry service. Nader et al. \cite{nader} also discuss a service oriented framework for managing smart-city based services through Fog and Cloud infrastructures. In this framework, services are classified in two types. The first one manages the core operations of the framework including resource management and security. and another type incorporates the requirements of specific applications. The security of the framework is ensured by authentication and access control mechanisms.
\par In the aforementioned frameworks, security issues are exploited from limited perspective. Furthermore, the computing capabilities of both edge and remote resources have not been fully leveraged. In some cases, pushing computation towards IoT devices or resource enriched Fog nodes increases overall deployment cost and energy consumption. In addition, most of the frameworks overlook the heterogeneity within the computing infrastructures. It is very difficult for them to support multiple applications simultaneously. However, our proposed FogBus combines the concept of prototype and platform-based frame-work so that it can offer service tuning facility to both users and service providers. It ensures data integrity through Blockchain and assists user authentication and data encryption side by side. FogBus expands the execution platform for different IoT applications from resource constrained Fog nodes to Cloud datacenters going beyond their diversity.  
\section{FogBus Framework} \label{mainPart}
\begin{figure*}[t]
\centering 
\includegraphics[width=175mm, height= 60mm]{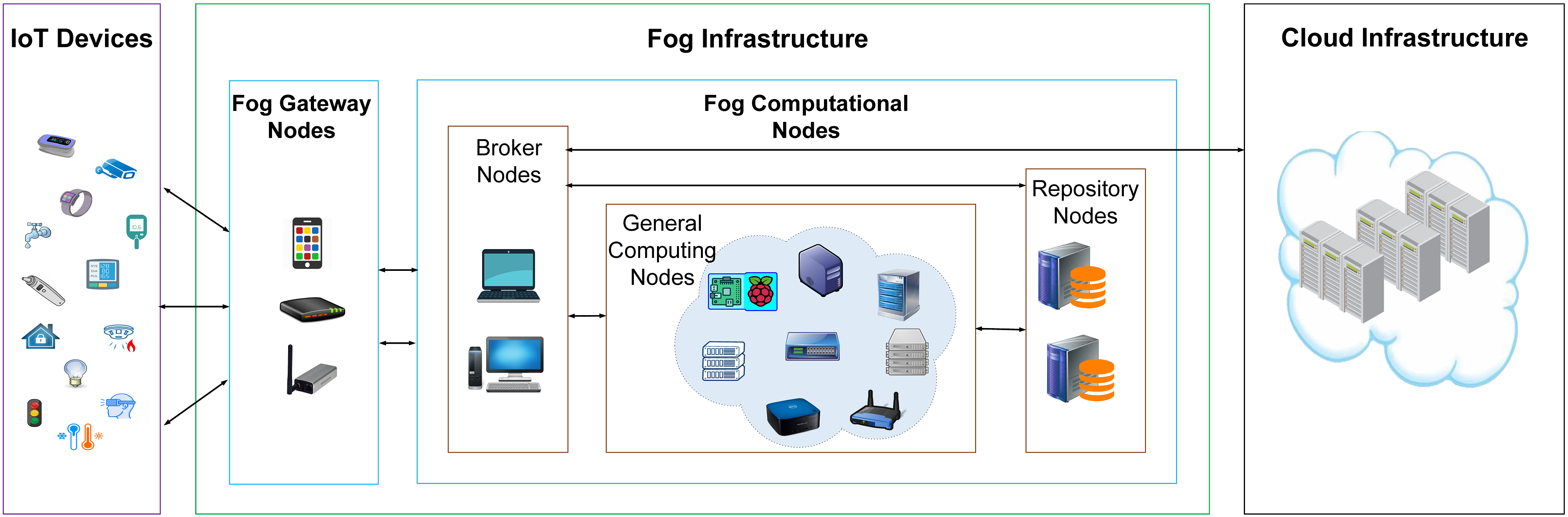}
\caption{High level view of integrating IoT-Fog and Cloud through FogBus framework}
\label{Fig:fogbus}
\end{figure*}

The FogBus framework integrates diverse hardware instruments through software components that offer structured communication and platform independent execution interfaces. High level view of integrated IoT-Fog-Cloud environment using FogBus is presented in Fig.\ref{Fig:fogbus}. The FogBus framework includes the following elements.    

\subsection{Hardware Instruments}
The hardware instruments that form the basis of FogBus such as IoT devices, Fog Gateway Nodes (FGN), Fog Computational Nodes (FCN) and Cloud datacenters are discussed below.     
\par \textbf{\textit{IoT devices}}: IoT devices contain sensors that perceive the external environment and actuators that convert any given command to physical actions. Usually, IoT devices are energy and resource constrained and act as mere data producer or consumer. In some cases, IoT devices are equipped with limited computation capabilities to preprocess raw data generated in real-time. FogBus allows IoT devices to connect with proximate gateways via wireless or wired communication protocols such as Zigbee, Bluetooth and NFC. The sensing frequency of IoT devices can be tuned according to context of the system where the format of IoT-data varies from device to device.     
\par \textbf{\textit{Fog Gateway Nodes (FGN)}}: In FogBus framework, Fog Gateway Nodes (FGN) are the entry points of distributed computing infrastructure. FGNs assist IoT devices to get configured with integrated environment for placing and executing corresponding applications. Through FGNs, the FogBus framework offers user interfaces of applications so that the users can set authentication credentials, access the backend program, convey service expectations, receive service outcome, manage IoT-devices and request resources from computing infrastructure according to their affordability. In addition, FGNs operate data filtration and organize them in a general format. FGNs also aggregate the data received from different sources of a particular smart system. For large scale processing of the data, FGNs forward them to other computing instances of integrated environment. In attaining this purpose, FGNs maintain rapid and dynamic communication with accessible Fog nodes by the use of either Constrained Application Protocol (CoAP) or Simple Network Management Protocol (SNMP) \cite{protocols}.   
\par \textbf{\textit{Fog Computational Nodes (FCN)}}: FogBus is designed to deal with numerous Fog Computational Nodes (FCNs) simultaneously. FCNs are heterogeneous in terms of capacity and resource architecture. They are equipped with processing cores, memory, storage and bandwidth to conduct various FogBus operations. Based on these operations, FCNs can act in three different roles:
\begin{enumerate}
\item \textit{Broker nodes}: To handle the back-end processing of IoT-applications, FogBus facilitates the corresponding FGNs to connect with any of the accessible FCNs. This FCN initiates back-end processing of the application provided that required resources for the operation are available within it. If it fails to meet application's requirements by itself, as a broker node it communicates with other FCNs and Cloud datacenters on behalf the FGN to provision required resources for executing the back-end application. In this case, it distributes the computational tasks over multiple FCNs and seamlessly monitors, synchronizes and coordinates their activities. FogBus supports such broker nodes with adequate security features and fault tolerant techniques such as Blockchain and replication so that they can ensure reliability in communication and exchanging data among FGNs, FCNs and Cloud datacenters.        
\item \textit{General Computing Nodes (GCNs)}: For security issues, FogBus does not expose all FCNs as directly accessible to the FGNs. They are used for general computing purposes and accessible via broker nodes. The broker nodes also explicitly manage their resources and forwards the data along with executable back-end applications for processing. A general computing node can simultaneously serve multiple broker nodes without degrading consistency of their individual operations. In addition, computing nodes form clusters among themselves under the supervision of specific broker node while executing distributed applications.           
\item \textit{Repository nodes}: Apart from conducting brokering and computing operations, some FCNs manage distributed database to facilitate data sharing, replication, recovery and secured storage. The repository nodes offer interfaces for instant access and analysis of historical data. They maintain meta-data of various applications including application model, runtime requirements and dependencies. Moreover, these nodes can preserve some intermediate data during application execution so that data processing can be started from any anomaly-driven stopping point.
\end{enumerate}
\par \textbf{\textit{Cloud datacenters}}: When Fog infrastructure becomes overloaded or service requirements are latency-tolerant, FogBus extends resources from Cloud datacenters to execute back-end IoT applications. Through Cloud datacenters, FogBus expands the computing platform for IoT applications across the globe. In association with Fog repository nodes, it facilitates extensive data storage and distribution so that access and processing of data become location independent.              
\subsection{Software components}
\begin{figure}[t]
\centering 
\includegraphics[width=90mm, height=65mm]{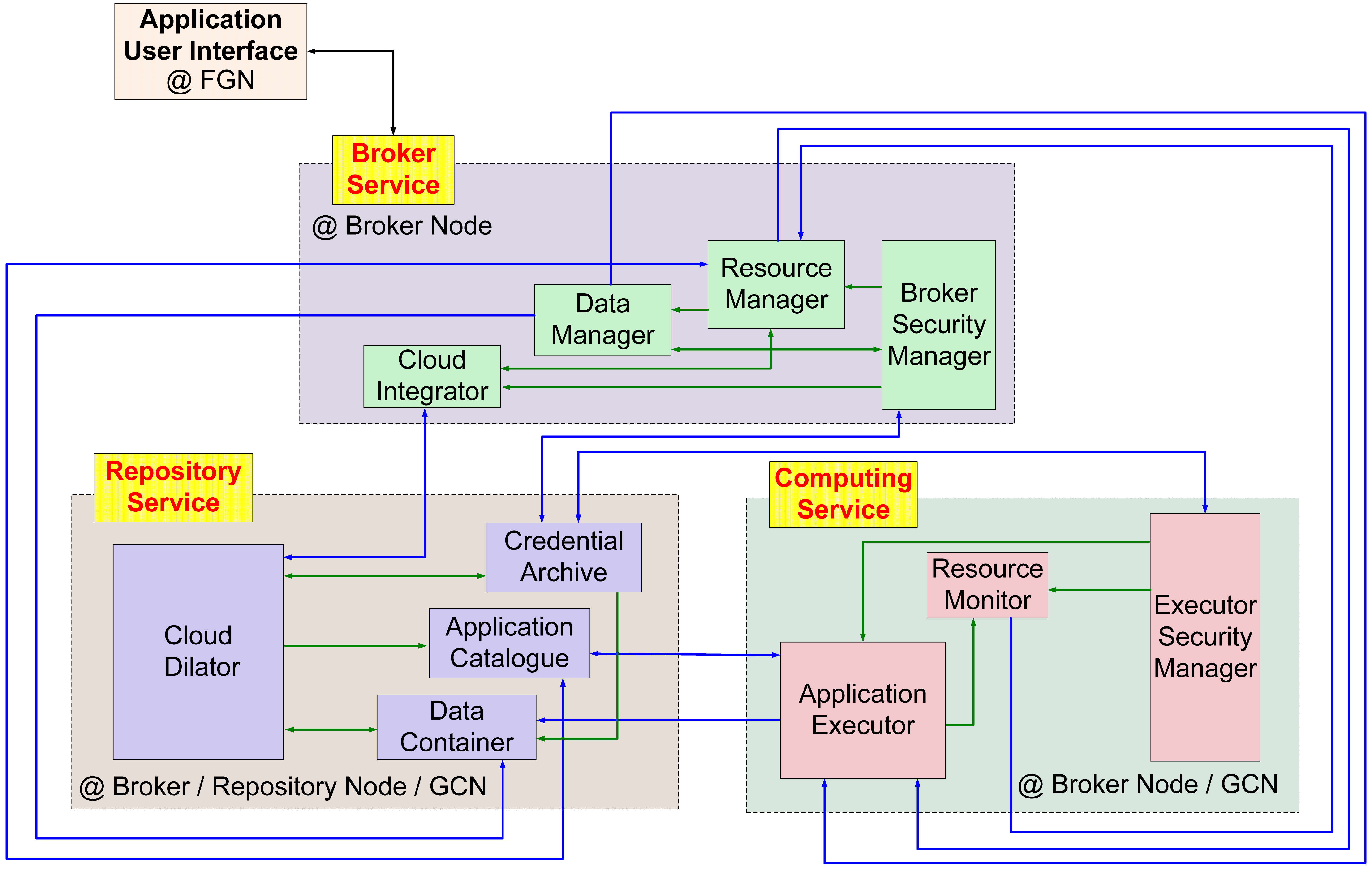}
\caption{Interaction of different software components within FogBus framework}
\label{Fig:software}
\end{figure}

To simplify IoT-Fog-Cloud integration, FogBus provides various interrelated and platform-independent software components. These components are broadly classified into three types of \textbf{System Services}. The Broker service manages all the functionalities of a broker node and initiates other software components according to the necessity, whereas the Computing service is responsible for controlling the operations of a general computing node. When a broker node itself starts the execution of back-end applications, the computing service is triggered within it. Conversely, Repository service can run across all the Fog nodes to mange repository-related operations. The interaction among different FogBus software components are presented in Fig. \ref{Fig:software}. The details of FogBus software components are discussed as follows.      
\subsubsection{Broker Service}
\textbf{\textit{Broker Security Manager}}: After receiving user's authentication credentials from particular FGN, the Broker Security Manager verifies them in association with Credential Archive of Repository Service. The Credential Archive also assist this component with required security certificates for remote Cloud integration. The Broker Security Manager itself generates the public and private key value pairs to facilitate port knocking, privileged port authentication and attribute-based encryption for securing the communication of corresponding broker node with other Fog nodes. Additionally, this component acts as the Blockchain interface for ensuring integrity while exchanging data with multiple entities. In this case, with the help of Data Manager it creates new blocks from the received data. The hash values and proof-of-work for each block are sent to Credential Archive for distributing among other nodes so that consistent verification of the chain can be ensured at different destinations. The Broker Security Manager along with Credential Archive and Executor Security Manager of Computing service manage further security issues within FogBus and offers other components flexible accesses to the required information.       
\par \textbf{\textit{Resource Manager}}: This component is responsible for selecting suitable resources to execute applications. It identifies the requirements of different applications from Application Catalogue of Repository service and perceives the resource status within each broker and general computing nodes through Resource Monitor of Computing service. The Cloud Integrator assists the Resource Manager with contextual data of Cloud-based instances such as virtual machines and containers. After attaining all information regarding resources and applications, Resource Manager provisions required resources on FCNs and Cloud for applications. In this case, Application Executor from Computing service and internal software system of FCNs and Cloud receptively helps the Resource Manager. Moreover, FogBus facilitates service providers to apply various policies in Resource Manager, while provisioning resources for applications. In addition, it maintains a resource configuration file that tracks the addresses of FCNs and Cloud instances along with deployed applications so that subsequent data of the streams can directly be sent to the allocated resources for processing. This file is also shared with Cloud for recovering the placement information during failure of the corresponding nodes. 
\par \textbf{\textit{Data Manager}}: This component receives the sensed and pre-processed data from the IoT devices. It can also aggregate data from multiple sources and calibrate data receiving frequency according to the context. However, with this data, blocks and their chains are created for maintaining integrity in association with the Broker Security Manager. Later it forwards the data to Application Executor of Computing service for processing and stores them in encrypted manner on Data Container of Repository service for further usage. After deployment of applications on allocated resources, Resource Manager shares the resource configuration file to Data Manager so that it can directly send subsequent data of the stream to the processing destination.
\par \textbf{\textit{Cloud Integrator}}: All interactions of FogBus framework with Cloud are handled by Cloud Integrator. It notifies the context of Cloud instances to the framework and forwards the storage and resource provisioning commands to the Cloud. Through this component, FogBus not only offers interface to the providers for developing customized Cloud-integration scripts but also facilitates access to third-party software systems to deal with multiple Cloud datacenters simultaneously. 
\subsubsection{Repository Service}
\par \textbf{\textit{Credential Archive}}: Users authentication credentials are set during IoT device configuration and are preserved in Credential archive. It distributes the security keys and details of each data block generated by the Broker Service to others.  This component also provides the Secure Socket Layer (SSL) and Transport Layer Security (TLS) certificates for Cloud integration. In addition, it supports Data Container for encrypting and decrypting the stored data. Through Cloud Dilator of Repository service, it periodically updates its image on Cloud so that security attributes can be recovered and distributed among others easily after uncertain failure of the corresponding nodes. 
\par \textbf{\textit{Application Catalogue}}: This component is responsible for maintaining the details about various types of applications including their operation, execution and programming model. Moreover, it specifies resource requirements and dependencies of the applications and their member tasks. The Application Catalogue can extend this information from Cloud through Cloud Dilator. Based on its provided specifications, Resource Manager of Broker service provisions resources for an application. According to the commands of Resource Manager, it also synchronizes the applications on allocated resources in association with the Application Executor of Computing service. 
\par \textbf{\textit{Data Container}}: Data received from IoT devices is stored in Data container so that it can be used for long term analysis. Here data privacy is ensured by applying encryption techniques. During application execution, it also receives some intermediate data from Application Executor that helps FogBus to restart the processing of data from any halting point. Moreover, in FogBus, the schema of Data Container based databases can be customized and shared according to the requirements of different IoT-enabled systems. In addition, Data Container maintains simultaneous association with Cloud Dilator to grasp the remote data and disperse the local data through Cloud.
\par \textbf{\textit{Cloud Dilator}}: This component facilitates other software components of Repository service to interact with Cloud. In this case, the Cloud Integrator of Broker service assists Cloud dilator with required commands for extending application specifications, transferring security attributes and exchanging data.                 
\subsubsection{Computing Services}
\par \textbf{\textit{Executor Security Manager}}: While conducting computing operations, the seamless secured interactions of an FCN with others are managed by the Executor Security Manager. In this case, the Credential Archive of Repository service assist this component with required security attributes. Along with Credential Archive and Broker Security Manager, this component plays a significant role in verifying the Blockchains.    
\par \textbf{\textit{Resource Monitor}}: Both busy and idle status of computing resources are monitored by this component in association with the Application Executor. These perceived information helps Resource Manager to provision resources for different applications. It also tracks the runtime QoS requirements of the application and the performance of the resources in meeting them. When allocated resources perform less than expectations, in dealing with the application or their uncertain failure occurs, this component immediately notifies the Resource Manager to initiate required actions such as dynamic resource provisioning, application execution migration and intermediate data storage.   
\par \textbf{\textit{Application Executor}}: Based on the provisioning instructions issued by the Resource Manager, this component allocates resources for different applications on corresponding FCN. It also extends the application executables from Application Catalogue to deployment on allocated resources. Once the application deployment is conducted, it begins to receive data forwarded by Data Manager for processing. In addition, this component periodically informs the status of resources to the Resource Monitor. When any anomaly is detected or predicted, this component is asked by the Resource Manager to extract intermediate data from application execution and store them on Data Container to make the framework fault tolerant.
\subsection{Network Structure}
\begin{figure}[!t]
\centering 
\includegraphics[width=88mm, height=60mm]{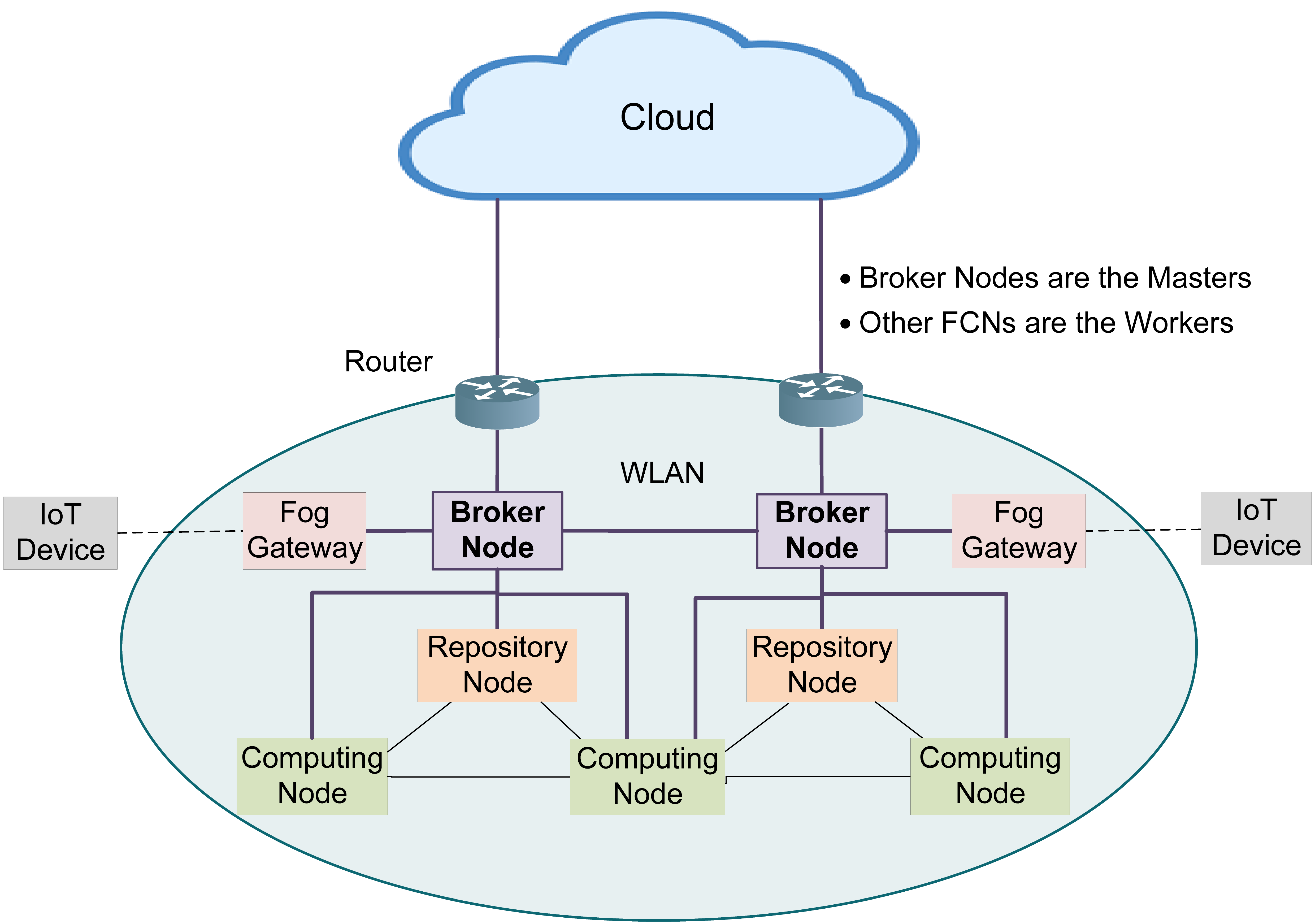}
\caption{Network Structure of FogBus framework}
\label{Fig:network}
\end{figure}

The software components of FogBus share numerous data and information among themselves. To facilitate their interplay, persistent and stable network communication among hardware instruments of the framework is necessary. It is also required to ensure that hardware instruments do not become overwhelmed with the communication burden. In addition, the FogBus networking should be secured, scalable and fault tolerant. Taking cognizance of these facts, we design the FogBus network structure as shown in Fig. \ref{Fig:network}. Different aspects of the network structure are described as follows. 
\par \textbf{\textit{Topology}}: The master-worker topology is applied in designing the network structure for FogBus framework. \textbf{Here, broker nodes are the masters while other FCNs function as the workers}. Being master, a broker node receives the data stream and user information from the FGNs and discovers workers for processing and storing them. During application runtime it manages functionalities of the workers and delivers the service result to FGNs derived from the application execution. In addition, it connects the Fog infrastructure of a FogBus enabled system with the Cloud infrastructure. To foster data sharing and reduce overhead from the masters, worker nodes also communicate among themselves under the explicit supervision of the masters. The masters, workers and FGNs of a FogBus-enabled system are connected with a common wireless local area network (WLAN) that is managed by one or multiple routers. 
\par \textbf{\textit{Scalability}}: FogBus framework allows service providers to scale-up the number of active Fog nodes according to context of the system. An FCN connected with the same WLAN can simply become a worker by making itself accessible to the corresponding master. Later, the master configures required software components on that FCN to conduct desired operations. The FogBus supports coexistence of multiple masters in a WLAN so that FGNs can get diverse options to dispatch the data streams for processing. The masters also share workers among themselves. In this case, the data integrity and privacy are not affected since each master maintains its own chain of blocks and separate database on the workers. In addition, the software components running at the masters facilitate Fog infrastructure to integrate with multiple Cloud datacenters simultaneously.
\par
\begin{figure}[!h]
\centering 
\includegraphics[width=80mm,height= 40mm]{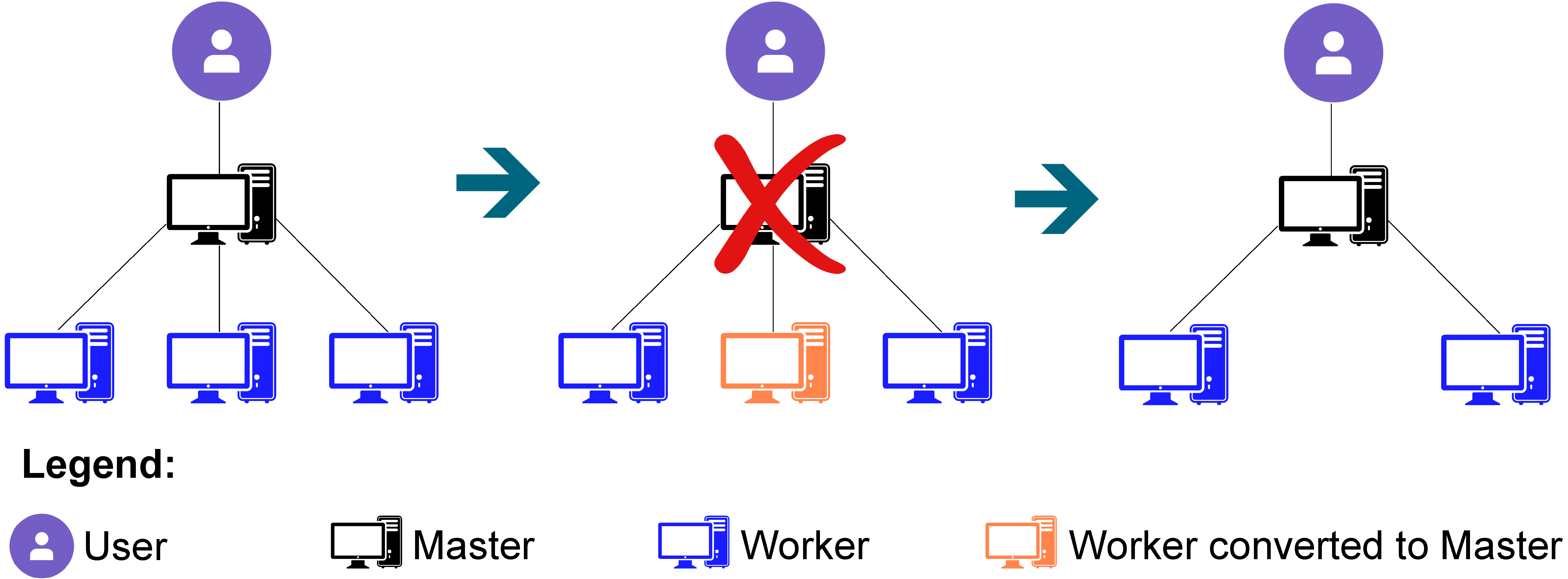}
\caption{Ensuring reliability in FogBus framework}
\label{Fig:reliability}
\end{figure}

 \textbf{\textit{Reliability}}: The facility of running multiple masters implicitly eradicates the inherent single point failure limitation of master-worker model within FogBus. Additionally, the framework allows each master to replicate their image over one of its worker nodes. During uncertain failure of that master, this replication operation helps corresponding worker to get the master privileges and defend the collapse of communication network as shown in Fig. \ref{Fig:reliability}. Here, the platform-independent characteristic of FogBus software components plays the key role. The masters also periodically check status of its workers and store their intermediate data and configurations including deployed applications in different places. When a worker fails, the masters share the worker's information with other workers so that residual data processing can start immediately. If all the workers of a master are overloaded, workers of other masters are taken into account. In this case, all the masters maintain an internal communication among themselves. Thus, the computation facility remains always available within the framework.
\par \textbf{\textit{Security}}: The inclusion of FGNs and FCNs in FogBus provided network require proper authentication. It is explicitly handled by the routers managing the WLAN. The masters also apply network level access control and packet filtration techniques to resist the network infrastructure from being compromised and eliminate the malicious contents. In FogBus, multiple communication links also exist to reach different Fog nodes. It eventually helps to readjust the routing path when any network anomaly is perceived. Cloud provided network security policies are further used in FogBus framework while interacting with different Cloud infrastructures. 
\par \textbf{\textit{Performance}}: FogBus framework utilizes the network bandwidth dedicatedly for a specific system. Since the network resources are not shared with external entities, the overall performance of the system from network perspective does not degrade. If the service providers intent to increase the number of FCNs in the framework, requirements for additional network resources will not be very high as well. In addition, it facilitates easy deployment of the Fog nodes and faster service delivery to the users. As a localized network, its throughput also remains at an acceptable level with the course of time. Additionally, FogBus supports periodic adjustment of network resources so that it can deal with any frequency and volume of incoming traffic. Moreover, the network structure of FogBus does not depend much on external hardware instruments for managing and configuring the network operations that implicitly reduce the capital and the operational expenditure for the service provider.
\section{Design and Implementation} \label{implementation}
The functionalities of FogBus encapsulated Application Programming Interfaces (API), execution environment, scripting and programming languages are supported by all hardware of the integrated environment. It eventually helps FogBus to function beyond the infrastructure heterogeneity. The implementation of different FogBus elements are described as follows. 
\subsection{System Services}
The System Services such as Broker service, Repository service and Computing service of FogBus are implemented as web programs. They are developed on PHP, an HTML-embedded server-side scripting language and use HTTP protocol based RESTful APIs to exchange data and share information among different FCNs within the WLAN. Usually these PHP based web programs can function in every operating system such as Unix, Windows, Linux and NetWare. On the other hand, most of the embedded, networking and IoT devices are either designed with built-in protocol stacks for HTTP communication or support their easy installation. Thus, the System Services of FogBus can run across different types of platforms. In FogBus, an Apache server is setup in each FCN to run the web programs of corresponding System Service. In addition, MySQL servers are installed in Broker and Repository nodes to manage databases and their operations. Each System Service of FogBus is divided into \textbf{Service Interface} and \textbf{Management Activity}. Apart from Blockchain, the Service Interface and Management Activity of each System Service handles other security aspects of the FogBus. Moreover, at masters, the Service Interface assists in receiving data and user's specifications from the gateway devices and presents the service results. Service providers also notify the workers IP addresses to masters through this interface. The Management Activity within the master contains the resource provisioning policies and updates the configuration files. Additionally, it forwards commands for the workers. The worker's Service Interface functions as a receptor of the corresponding node and is responsible to decode the output file of applications to masters. Based on the master's signals, the Management Activity at worker functions monitoring of the resources, circulating their status to the masters and allocation of resources. It also stores the received data in relational databases and creates input file for backend program of applications.        
\subsection{Blockchain}
Maintaining integrity of data and ensuring that data is not sent by an unregistered source are very important for credibility of the system. For data integrity and data prevention from tampering, Blockchain technology is recently adopted in many real-time systems \cite{blockchain}. Theoretically, Blockchain is a set of distributed ledgers that can be programmed to record and track the value of anything. In Blockchain, whenever new data is received by an entity of the distributed system, it forms the data into a block. This block possesses a hash value that is usually created by using the corresponding data, index of the block in the chain, the timestamp of the data reception and the hash of its previous block within the chain. Additionally, the node mines the block with other blocks of the chain to create a proof-of-work for that block so that its hash follows a similar pattern with others. Later, the data, copy of the block is sent to other nodes for linking with their local chains. In this operation, nodes mine the block to certify the proof-of-work. Digital signature is also used to veryfy source of the block at the destination. However, if the data of any block is altered on a node, the hash of that block will change and mismatch with its hash saved in the next block. As a consequence, the later part of the chain will become invalid. To make the chain valid again, hash of the invalid blocks are required to be recalculated. Besides, the proof-of-work of each block requires to be generated again. Both of these operations are time consuming and compute intensive. Moreover, this fraudulent manipulation of data in a Blockchain will not be successful unless 50\% of its distributed copies are individually reformed by following the same set of operations. Thus, it becomes very hard to alter any data in Blockchain within rigid time limit \cite{blockchainbook}. 
\par In FogBus, the masters create the blocks from received data and calculate the hash of each block based on the data, hash of the previous block, time stamp and a nonce value using SHA256 algorithm \cite{mitBlock}. Masters also create random public/private key pairs that help to generate unique signatures with the original data. Later they share Blockchain details, digital signature attributes and the data in Base64 encoding format with workers. With the received public key of the masters, the workers are able to verify that the data is coming from a legitimate source. If any other key is used, that data is rejected. The public-private key pair in this case is kept dynamic per block to prevent the generation of private key using brute force techniques. Additionally, each block and its hash are verified at the workers by mining the nonce value that supports the proof-of-work. If any worker reports error in terms of Blockchain tampering or signature forgery, then the Blockchain in majority of the network is copied to that node. FogBus also offers users and service providers to track the data/block flow through the Service Interface running at masters by displaying the latest hashes of the Blockchain copy at each worker. Thus, it helps users and service providers to take necessary action on suspicious activity within the FogBus network. In FogBus, the Blockchain is developed in Java programming language. Compiled Java programs usually run on Java Virtual Machine (JVM), which can be easily installed across various platforms. Hence, the Blockchain utility of FogBus can function in wide range of operating systems. In different FCNs of FogBus, this utility directly interacts with the corresponding System Service.   
\par 
\subsection{Cloud Plugin}
In FogBus, the Service Interface running at master prompts the user to specify their intention regarding Cloud integration for data processing. If users wish to extend Cloud resources for computation, only then the Cloud Plugin of FogBus which is deployed on the master, becomes activated. For other operations such as storage and distribution, the Management Activity at masters directly communicates with the Cloud. However, FogBus offers flexibility to providers for using different customized or third-party Cloud Plugin services to integrate Cloud and Fog infrastructure for computing purposes. In the case of running third-party Cloud Plugin services, the master is required to configure according to the requirements of that plugin. However, to develop customized plugin, it is preferable to use cross-platform programming languages. In the current version of FogBus, Aneka, a third-party software is used for Cloud integration to perform computational operations.
\begin{figure*}[ht]
\centering 
\includegraphics[width=175mm,height=80mm]{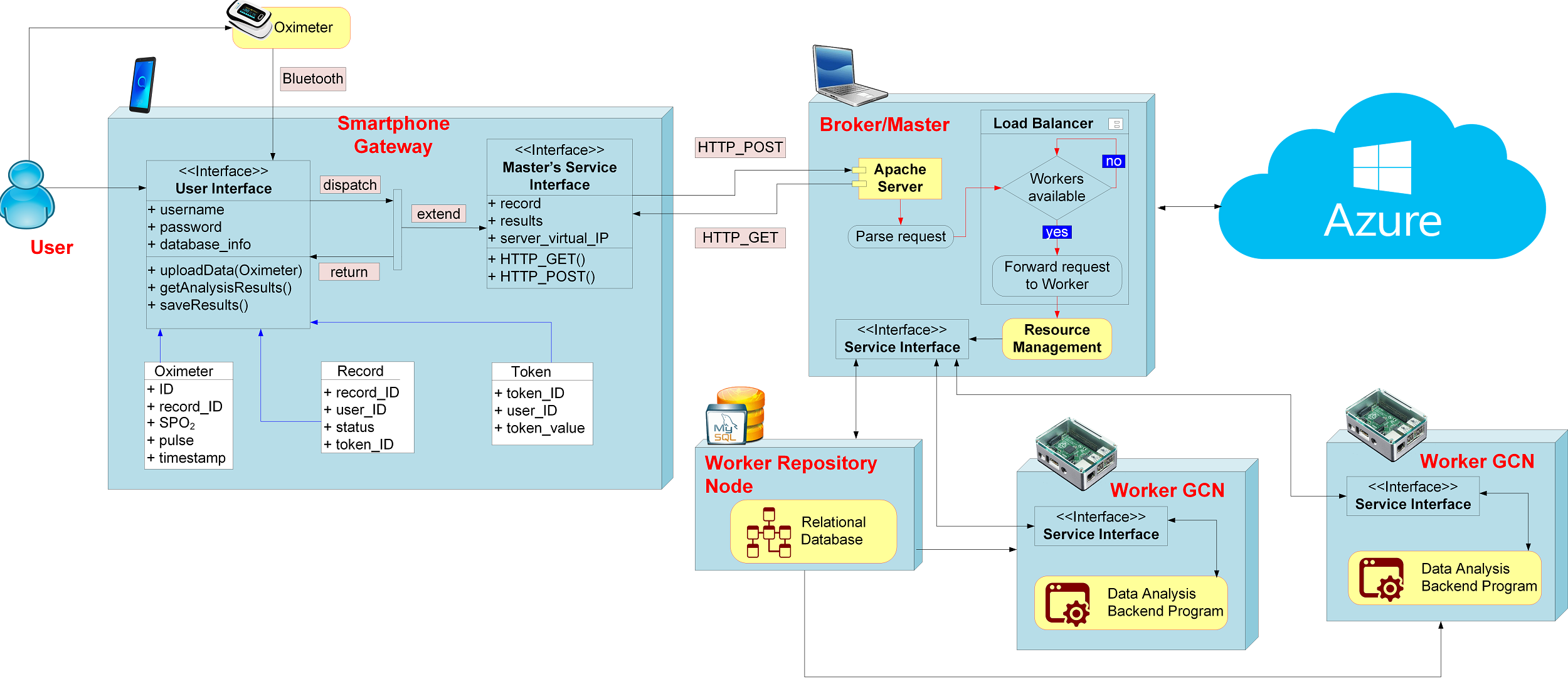}
\caption{FogBus framework enabled system model for Sleep Apnea analysis}
\label{Fig:pro_model}
\end{figure*}
\par Aneka is a PaaS framework for facilitating the management of Cloud-based applications \cite{aneka}. The Aneka framework functions in a service-oriented manner. It is equipped with a set of software components to configure, operate, and monitor an Aneka-Cloud environment. The Aneka-Cloud can be formed with heterogeneous instances from either public or private, or hybrid Cloud. Aneka offers the developers diverse APIs for provisioning and scheduling both physical and virtual resources in the Aneka-Cloud. Developers formulate the logic of applications using different programming models and set the runtime environments for their deployment and execution. Currently Aneka platform supports the Bag of tasks, Distributed threads, MapReduce and Parameter sweep model. In the Aneka-based Cloud plugin of FogBus, IP addresses of Cloud instances are specified by the providers. This plugin can initiate both task and thread model in Aneka-Cloud to conduct data processing on single and multiple Cloud instances respectively \cite{aneka-2}.  
\par According to the built-in resource provisioning policy of FogBus, at first Fog infrastructure is exploited to process data, later Cloud infrastructure is referred. For the second case in FogBus, the Management Activity at a master stores the data in a Cloud input file. The Aneka-based Cloud plugin at the master parses this file in every 500 milliseconds of polling period and checks for the pending data for processing. If any pending data exists, it forms either a task or threads; encapsulating the data at Aneka-Cloud and launches to one or multiple Cloud instances. In this case, Blockchain is also applied to ensure data integrity.
\subsection{Application}
FogBus framework supports the execution and deployment of applications of different IoT-enabled systems. In FogBus these applications are divided into user interface and backend program. Although applications are not the part of FogBus software components, FogBus offers developers some guidelines to build their user interface and back-end programs aligned with the features of FogBus framework. The required specifications of user interface and back-end program of applications are described as follows.  
\subsubsection{User Interface}
The user interface of applications runs in FGNs. The underlying platform of most of the FGNs are Android, iOS, Windows, Tizen, WebOS and RTOS. In this case, the programming language for developing the user interface should be supported by these platforms. Moreover, for some applications, user interface requires to store data temporarily. On that note, the developers should use compatible database system and schema to these platforms. Besides, the user interface deals with the incoming data from IoT devices. The majority of IoT devices run Bluetooth Low Energy network technology for communication as they are energy constraint. To handle this issue, the user interface should support both general and low energy Bluetooth interactions. In FogBus framework, user interface is directly correlated with the Service Interface running at the masters for forwarding IoT data and user information, and receiving the service outcome. For simplicity of the interaction, the user interface can be designed in such a way that easily parses the web programs of master's Service Interface. 
\subsubsection{Backend Program}
In FogBus, the backend program of applications is executed in the FCNs. Since the FCNs are distributed, to fully leverage their capabilities it is preferable to build the backend program in distributed manner. In this case, modular development of backend program can be applied by the developers. In addition, the execution of backend program should not be obstructed by the heterogeneity of FCNs. To address this issue, developers can use cross platform programming languages to develop the backend program. While developing the backend program some specific points within the script should be specified so that application's intermediate data on those points can be stored during execution. Furthermore, the backend program should be able to extract the input file and update the output file at the workers.       
\section{A Case Study : Sleep Apnea Analysis} \label{caseStudy}
In this work, FogBus framework has been adopted for deploying and executing a real-world application named \textit{Sleep Apnea Analysis}. Sleep Apnea is a disease in which air stops flowing into the lungs for 10 seconds or even longer period of time during sleep. Hence, it reduces oxygen level in blood of the patient, downs the heartbeat rate and resembles that the patient has stopped breathing. It can happen very frequently and create severe obstruction in sound-sleep of the patient. Furthermore, if oxygen saturation becomes significantly low for aged and asthma patients, Sleep Apnea could provoke cardiac failure or brain stroke. However, Sleep Apnea is a very common disease although most of the people either ignore or unaware of it. To determine the intensity of Sleep Apnea, it is required to monitor oxygen saturation rate in blood time to time. If the intensity becomes higher than normal, it is recommended to consult with the Doctor before it occurs other complications \cite{sleep}.
\par Usually, Sleep Apnea analysis is difficult and cumbersome since it requires an overnight sleep study to grasp the necessary data. In this procedure, pulse oximeter and Electrocardiogram (ECG) machines are hooked up with various parts of the patient's body during sleep time. Based on the received peripheral capillary oxygen saturation, SpO2 and ECG data, the doctors determine Apnea Hypopnea Index (AHI) of the patients that presents the Sleep Apnea intensity in proportional manner. Currently, to conduct the Sleep Apnea analysis, hospital or laboratory-based machineries are required which are expensive to own by individuals. Besides, this analysis becomes very latency sensitive while critical patients are being monitored. Therefore, we develop a prototype for low cost Sleep apnea analysis using FogBus framework that gathers both SpO2 and heart beat rate from a finger pulse oximeter \cite{pulse} and harness local resources for their processing. It is affordable for patients, easily configurable and provides faster results compared to Cloud-based processing. The detail of FogBus-enabled Sleep Apnea analysis prototype is described as follows. 
\subsection{System Configuration} \label{model}
The system setup for FogBus-enabled Sleep Apnea analysis prototype is presented in Fig. \ref{Fig:pro_model}. The configuration of different hardware instruments are given below.   
\par \textbf{\textit{IoT Device}}: Jumper JPD-500F Finger Pulse Oximeter, 1.5V, Bluetooth Low Energy v4.2 (BLE), UTF-8 data encoding. 
\par \textbf{\textit{Gateway}}: Smartphone, Oppo A73 CPH1725, Android 7.1.1. 
\par \textbf{\textit{Broker/Master Node}}: Dell Latitude D630 Laptop, \textit{Intel(R) Core(TM)2 Duo CPU
E6550 @ 2.33GHz 2GB DDR2 RAM}, 32-bit, Windows 7, Apache HTTP Server 2.4.34, Java SE Runtime Environment (JRE) 1.6, MySQL 5.6, .net 3.5, Aneka 3.1.   
\par \textbf{\textit{Other FCN/Worker Node}}: Raspberry Pi 3 B+, \textit{ARM Cortex-A53 quad-core SoC CPU @ 1.4GHz 1GB LPDDR2 SDRAM}, IEEE 802.11, 64-bit, Raspbian Stretch, Apache HTTP Server 2.4.34, JRE 1.6, MySQL 5.6. 
\par \textbf{\textit{Public Cloud}}: Microsoft Azure B1s Machine, 1vCPU, 1GB RAM, 2GB SSD, Windows Server 2016, .NET 3.5, Aneka 3.1.
\par
\begin{figure}[!t]
\centering 
\includegraphics[width=90mm, height=50mm]{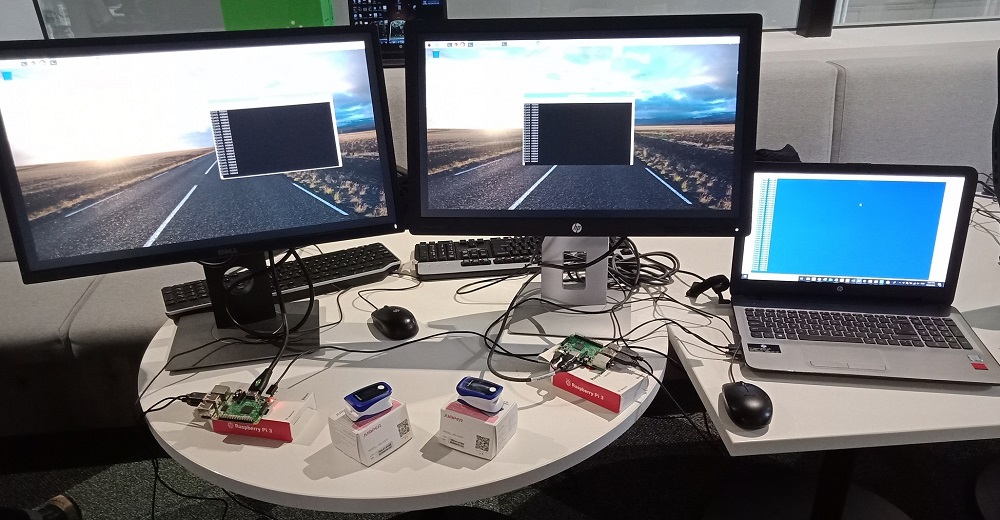}
\caption{Real-world implementation of FogBus-based Sleep Apnea analysis}
\label{Fig.real}
\end{figure}

 Fig. \ref{Fig.real} depicts the real implementation of this system model.
\begin{figure}[!t]
	\begin{center}
	\centering
		\includegraphics[width=25mm,height=40mm]{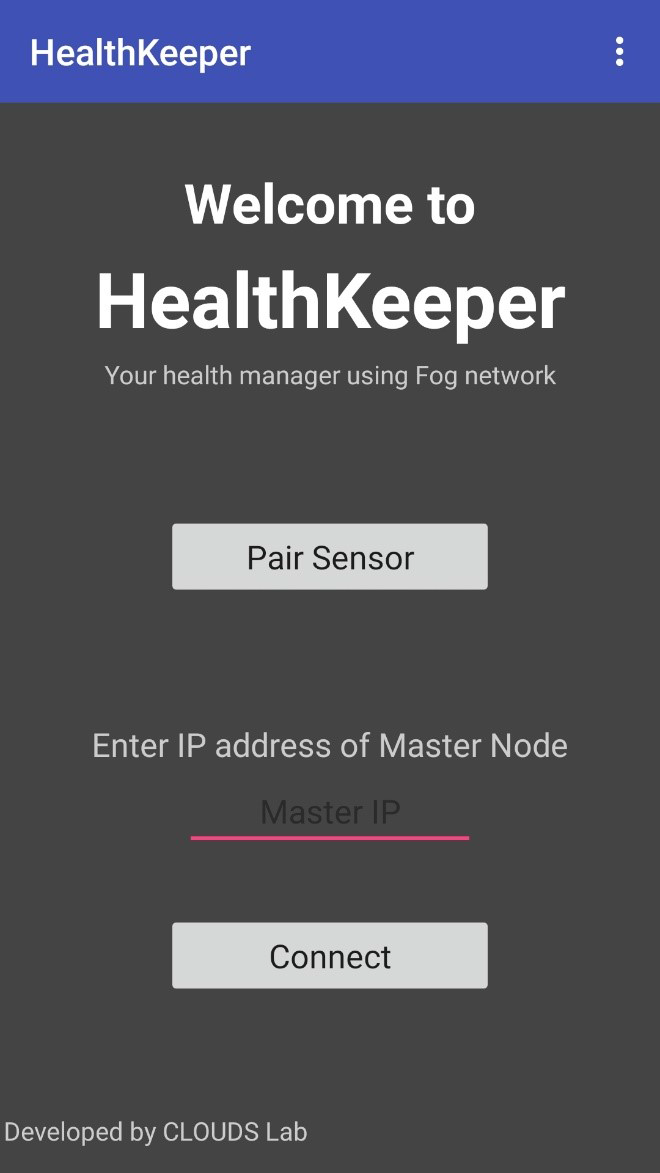}
		\hspace{20pt}
		\includegraphics[width=25mm,height=40mm]{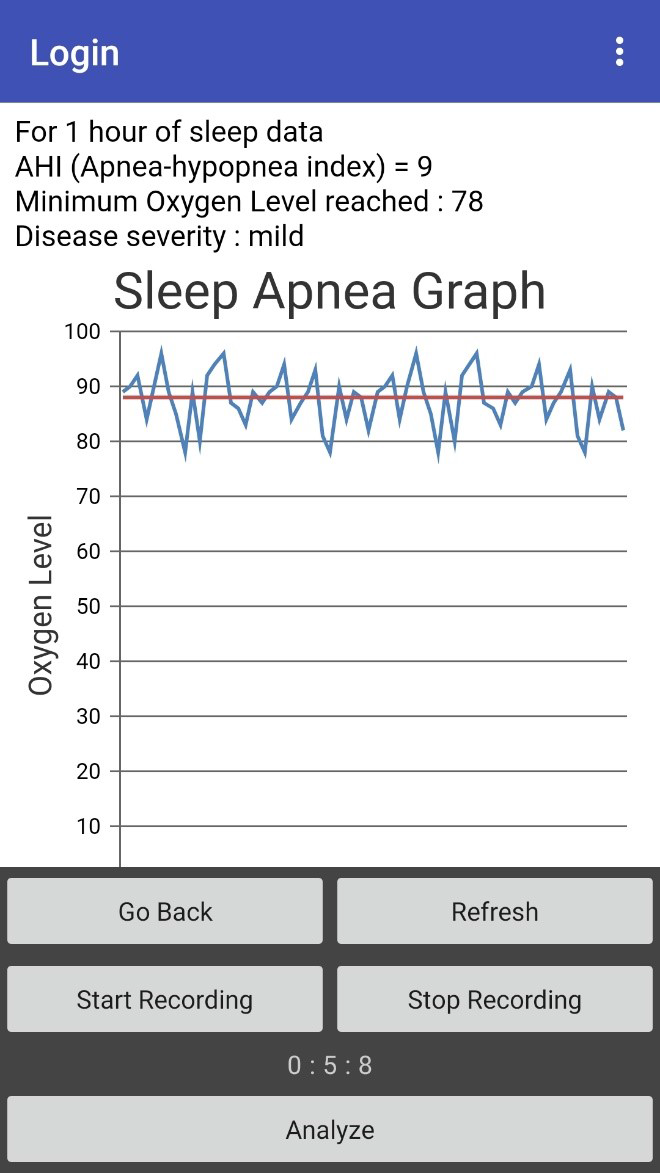}\\
        \footnotesize{(a)\hspace{120pt}(b)}\\		 
   	\caption{(a) Home and (b) Session Screen of the android interface} \label{fig:android}
	\end{center}
\end{figure}
\subsection{Installed Package}
The developed prototype for Sleep Apnea analysis is mostly Fog infrastructure centric. However, if Fog infrastructure is unable to process the data, using built-in Aneka-based Cloud Plugin of FogBus, the data is sent to Azure VM. The application package for Sleep Apnea analysis installed in the prototype consists of an android user interface and a data analytic backend program. Description of the installed package is given below.     
\subsubsection{Android Interface at Smart Phone Gateway}
An android executable named \textit{HealthKeeper} launches the android interface to the prototype operator. The executable installed on the Smartphone allows the device to act as an mediator between the Pulse Oximeter and the Master. It is developed on \textit{MIT App Inventor}, an open source platform \cite{Mit}. The interface is divided into \textit{Home} and \textit{Session} screen (Fig. \ref{fig:android}). The Home screen helps operator to pair the Oximeter with the Smartphone for receiving patient data using Bluetooth and enter the master's IP address. The Session screen handles all interaction with the master including data transmission. In this case, rather than sending data manually through the HTML form, the interface records and transmits data automatically. An empty data list is initialized and timer is reset when recording starts. Each data value received from the Oximeter is appended to the list. When the recording is stopped, the list is sent to the master for storage and distribution to the workers. This screen also extends the Service Interface running at the master and displays the result to operators once they become available to the master.   
\subsubsection{Data Analytic at Worker Computing Nodes}
The data analytic for Sleep Apnea analysis encapsulates two open source programs found in \cite{link1}\cite{link2}. These Java programs are stored in the repository worker and based on the command of master, they are forwarded to the computing workers for installation. The data analytic takes the input data as a file. From the input file the first, second and third columns are parsed as the timestamp, heart beat rate and blood oxygen level respectively. In the analytic, a Boolean variable tracks whether there is a dip in oxygen level or not. Whenever the oxygen level goes below 88, the dip Boolean variable turns to true and stays true till oxygen level is above 88. It is verified by the rise of heart beat rate in nearby timestamps of the dip occurrence in oxygen level. A counter variable in the analytic narrates how many times the dip Boolean variable has been changed to true. This count is known as the \textit{Apnea - Hypopnea Index, AHI} that is used to determine the intensity of Sleep Apnea. AHI based cases for Sleep Apnea analysis are given below.  
 \[
    \text{Sleep Apnea} =
    \left\{
    \begin{array}{lr}
      \text{No/Minimal},& \for AHI < 5 \text{ per hour}\\
      \text{Mild},& \for 5 \geq AHI < 15 \text{ per hour}\\
      \text{Moderate},& \for 15 \geq AHI < 30 \text{ per hour}\\
      \text{Severe},& \for AHI \geq 30 \text{ per hour}\\
    \end{array}
    \right\}.
  \]
However, as additional information, the data analytic stores the minimum oxygen level for the given period of time. For the heart rate data, minimum and maximum value are identified. The average heart rate and average rise or fall of the heart rates are also determined. In addition, heart beat pattern during the dips in oxygen level are filtered and ECG is generated. After identifying these information and Sleep Apnea intensity, the analytic delivers the result in a file. This file is later parsed by the master's Service Interface to notify the prototype operator.
\subsection{Sequence of Communication}
\begin{figure}[!t]
\centering 
\includegraphics[width=92mm, height=60mm]{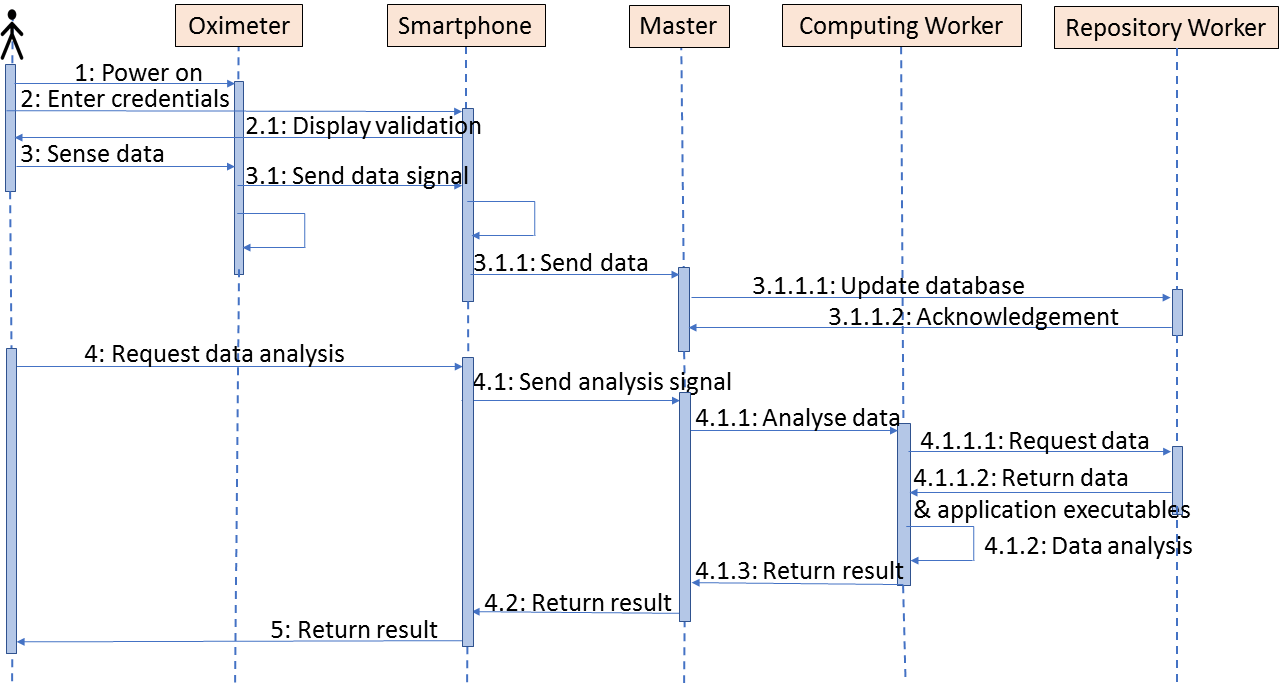}
\caption{Sequence of communication during Sleep Apnea analysis}
\label{Fig:pro_com}
\end{figure}

In the prototype of FogBus-enabled Sleep Apnea analysis, all hardware instruments belongs to same WLAN. Their sequence of communication is presented in Fig. \ref{Fig:pro_com}. This sequence initiates by configuring the Pulse Oximeter with the Smartphone using required credentials of the operator. The Oximeter senses patient's SpO2 and heart beat rate and forwards to the Smartphone through bluetooth communication. From Smartphone, these data are sent to the master. The master later stores the data on repository worker. After the storage operation acknowledgement is confirmed from the repository worker to the master. Since the Smartphone extends master's Service Interface, this acknowledgement becomes visible to the operator. 
\par After recording the data for a certain period of time, the operator prompts a request to the master via Smartphone for analyzing the stored data. Then, the master communicates with a suitable computing worker and issues required privileges for data analysis to it. The computing worker requests the stored data and analytic executable from the repository worker. On reception of these elements, the computing worker starts the analysis operation. Once the analysis operation is finished, the result is sent back to the master. The Smartphone pulls the result from the master and displays to the operator.                 
\section{Performance Evaluation} \label{performance}
\subsection{Experimental Setup}
The prototype for Sleep Apnea analysis discussed in Section \ref{caseStudy} is used to evaluate the performance of FogBus in terms of latency, energy, processor (CPU), memory (RAM), storage (Cache) and network usage. For the experiment, data from multiple pulse oximeters are recorded for a specific period of time, later the master sequentially generates analysis tasks for each recorded data chunk to the computing worker. Here, each experiment scenario is modelled under the following settings.    
\begin{enumerate}
\item \textit{With / Without Interval}: In With Interval setting, master sends the next analysis task to its computing worker after 5 seconds of receiving the outcome for previous task. This time interval helps both master and computing worker to reduce their overhead. On the contrary, in Without Interval case, master sends the next task to its computing worker as soon as the outcome of the previous task becomes available. It ensures that the FogBus framework remains consistently active and there exists no idle time on the nodes. 
\item  \textit{With / Without Blockchain}: FogBus offers flexibility to either enable or disable its Blockchain security feature according to the requirements of the users and service providers. The segments of this experiment setting differs from each other based on the status of Blockchain security feature in the FogBus-based prototype. 
\item \textit{Fog / Cloud Only / Integrated}: FogBus supports application execution across diverse computing infrastructures. This experiment setting refers whether the application execution is solely conducted on Fog or Cloud, or integrated infrastructure. 
\end{enumerate}
During the experiments, data parameters are recorded using Microsoft Performance Monitor at the Master and the Azure VM whereas at the Raspberry Pi circuits NMON Performance Monitor is used \cite{windows} \cite{linux}. Apart from the system model parameters specified in Section \ref{model}, additional parameters used for the experiments are given in Table \ref{Tab.sim}.    
\begin{table}[!h]
\centering
\scriptsize
\caption{Experiment parameters}\label{Tab.sim}
\begin{tabular}{|p{4.5cm}|p{2cm}|}
\hline
    Parameter & Value \\ \hline
    Analysis Task: & \\
    Duration of sequential task generation & 5 minute \\ 
    Data recording time per task & 3 minute \\ \hline  
    Pulse Oximeter: & \\
    Signal length & 18 KB \\
    Sensing frequency & 2 signal per second \\ \hline 
    WLAN: & \\
    Download Speed & 7 Mbps \\
    Upload Speed & 2 Mbps \\ \hline 
\end{tabular}
\end{table}
\subsection{Result Analysis}
\subsubsection{Number of Tasks}
\begin{figure}[h]
	\begin{center}
	\centering
		\includegraphics[width=35mm, height=38mm]{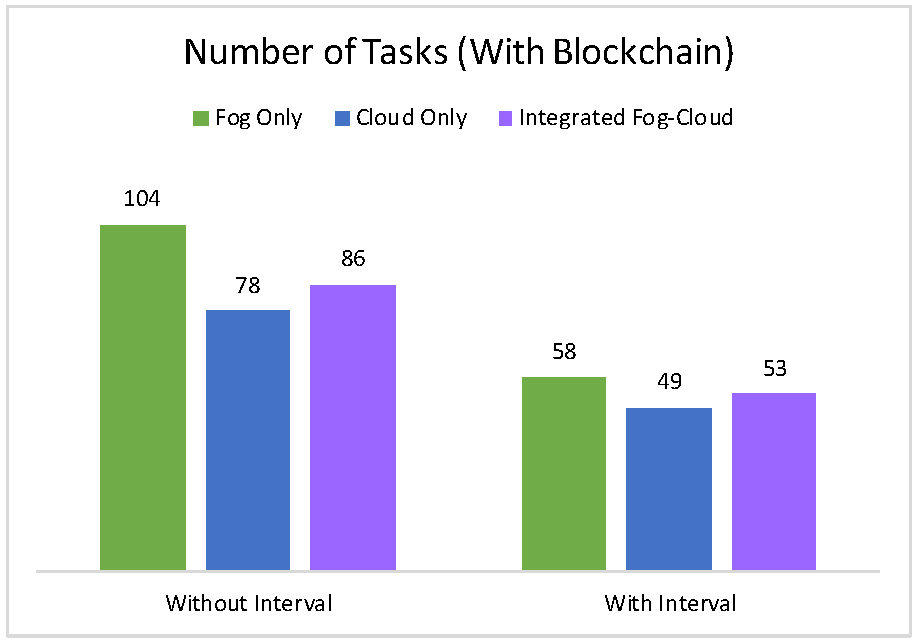}
		\hspace{20pt}
		\includegraphics[width=35mm, height=38mm]{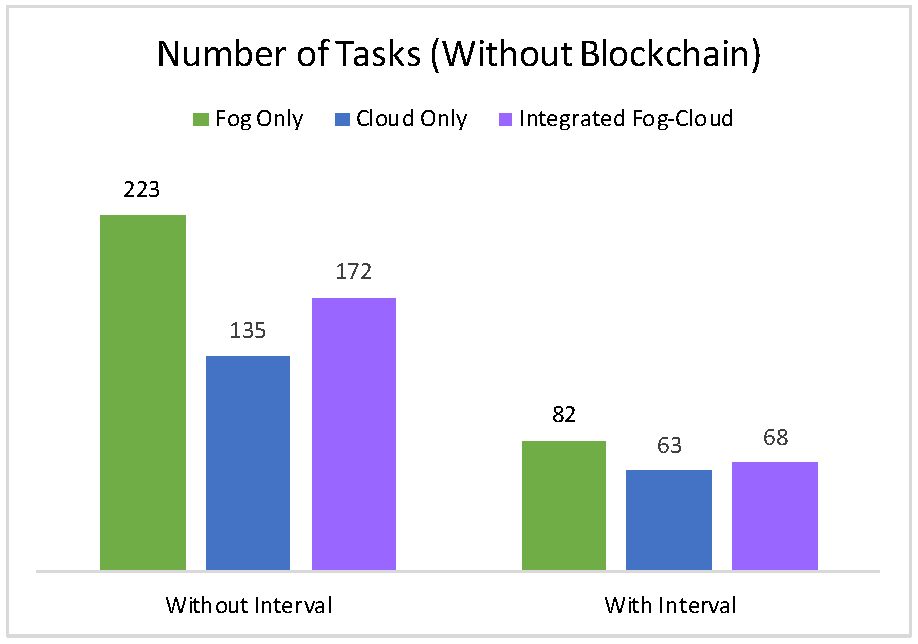}\\
        \footnotesize{(a)\hspace{120pt}(b)}\\		 
   	\caption{Number of tasks (a) With and (b) Without Blockchain} \label{fig:task}
	\end{center}
\end{figure}

Fig. \ref{fig:task} depicts the number of tasks generated in FogBus on different experiment settings. It is observed that the number of tasks is higher in the Fog Only setting compared to the Cloud Only and Integrated Fog-Cloud case. It happens since Fog infrastructure quickly delivers outcome of the previous task. During Without Interval setting, this value rises significantly than the With Interval setting since tasks are generated continuously by the master. It is also noticed that, if Blockchain feature of FogBus is turned off, comparatively higher number of tasks are generated. In this case, as lower amount of additional data is shared and processed over the infrastructure, it consequently improves the speed of receiving outcome for the previous task. Based on these observations, it is understood that if there exists higher amount of tasks to be handled with less security requirements, FogBus can be set to Fog Only setting with disabled Blockchain feature. However, in such state the management and processing overhead of the infrastructures will increase in proportion to the number of tasks and the size of data chunk for individual task. It can be managed by tuning the interval between subsequent tasks creation.            
\subsubsection{Latency}
\begin{figure}[h]
	\begin{center}
	\centering
		\includegraphics[width=35mm, height=38mm]{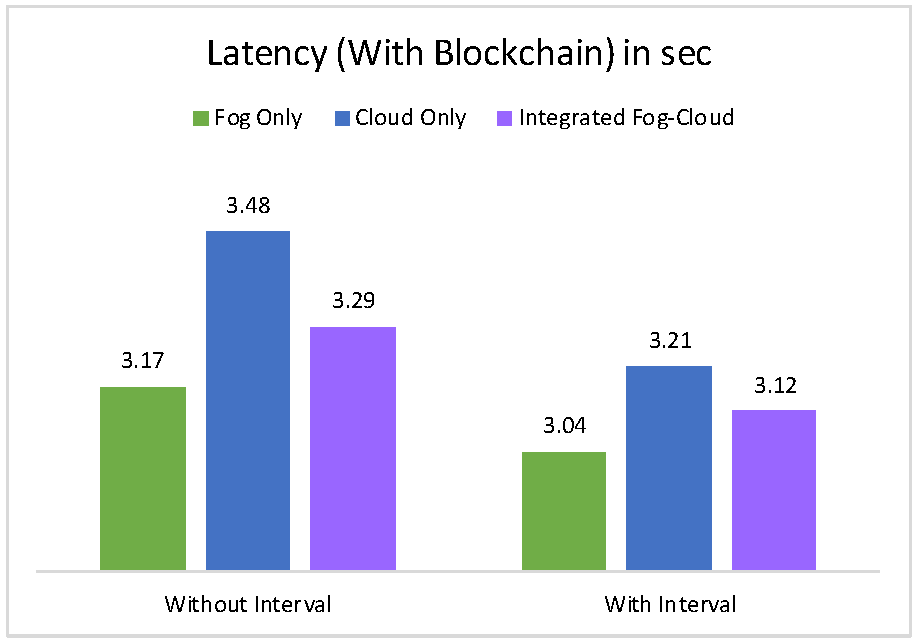}
		\hspace{20pt}
		\includegraphics[width=35mm, height=38mm]{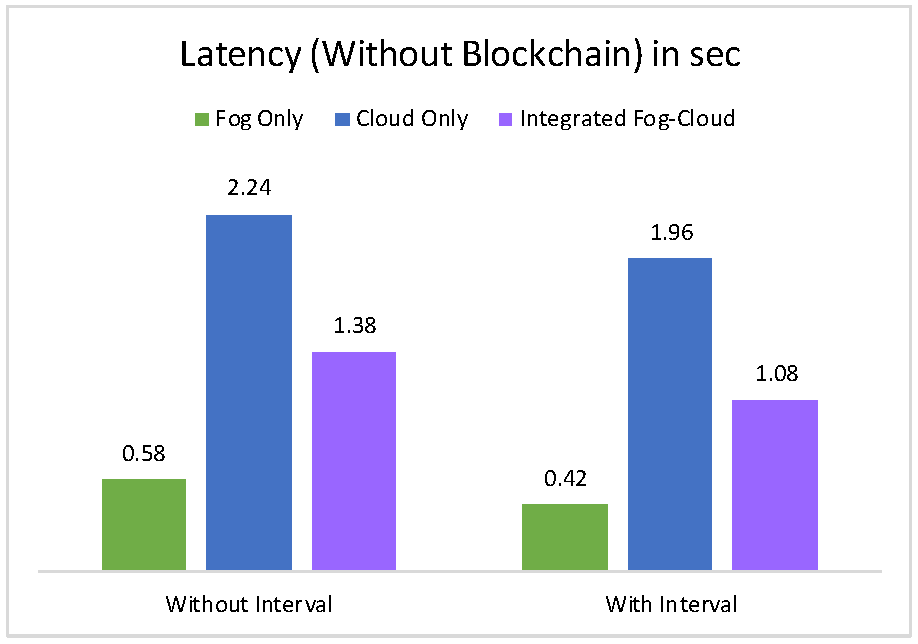}\\
        \footnotesize{(a)\hspace{120pt}(b)}\\		 
   	\caption{Latency (a) With and (b) Without Blockchain} \label{fig:latency}
	\end{center}
\end{figure}

Fig. \ref{fig:latency} presents the impact of different settings of FogBus on service delivery latency. Here service delivery latency is modelled as the summation of network propagation delay and task completion or application execution time. It is known that computational capability of Fog infrastructure is not enriched but it resides closer to the data source. As a consequence, network propagation delay is quite less for Fog infrastructure. Furthermore, if the size of data chunk for a particular task is not huge, its completion time will not differ significantly whether the application is executed in Fog or Cloud. Since, in this experiment, size of data chunk for a task is not huge, the service delivery latency much depends on the network propagation delay. As a result, in Fog Only setting of the FogBus, service delivery latency is minimal compared to Cloud Only and Integrated Fog-Cloud case. This latency becomes much lower on disabled state of Blockchain feature since its management add some more time to complete the tasks. Moreover, the With Interval setting reduces overhead from the infrastructure and network in this case; that also contributes to improve the service delivery latency. Therefore, it can be realized that these settings assist FogBus to deal with the tasks having stringent deadline.       
\subsubsection{Network Usage}
\begin{figure}[h]
	\begin{center}
	\centering
		\includegraphics[width=35mm, height=38mm]{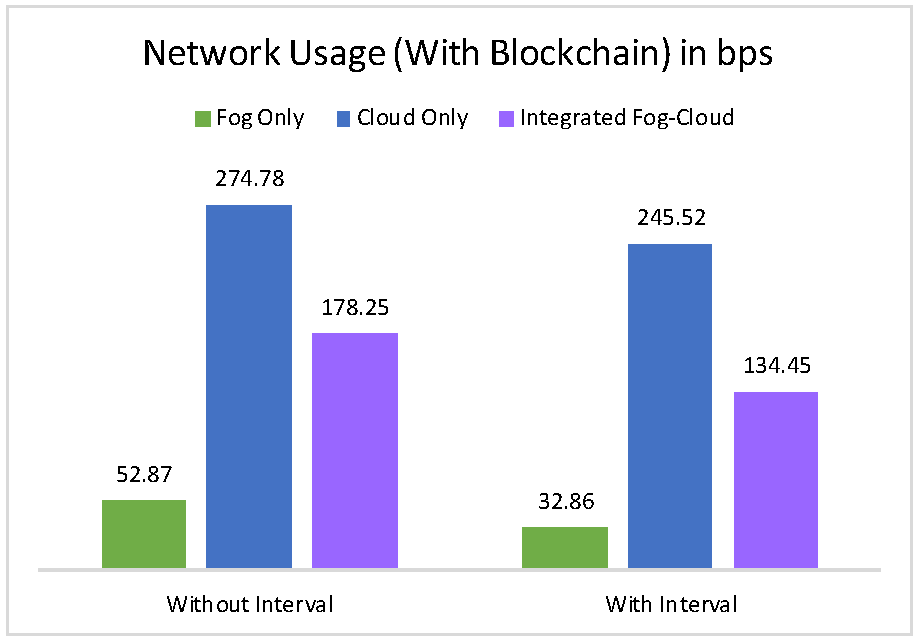}
		\hspace{20pt}
		\includegraphics[width=35mm, height=38mm]{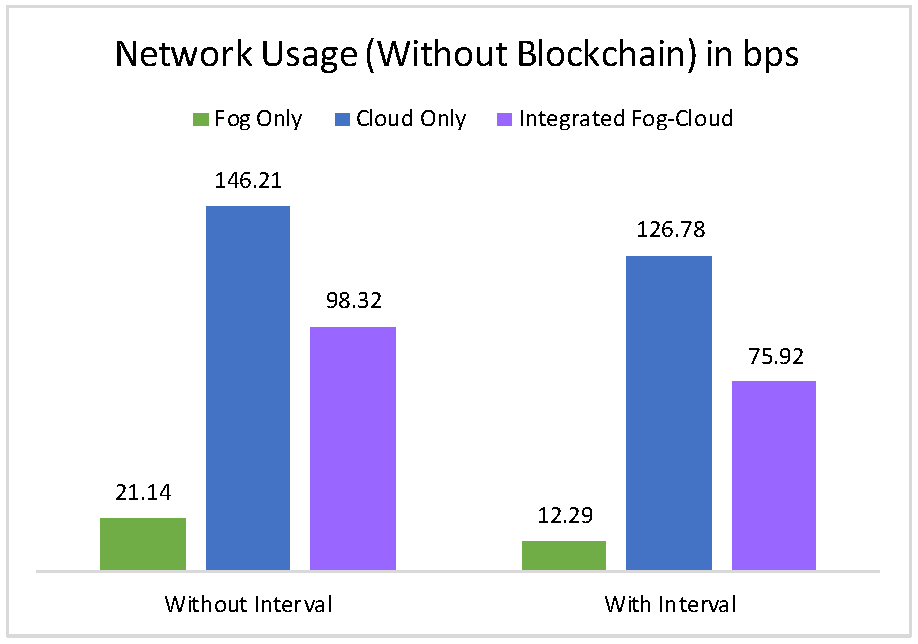}\\
        \footnotesize{(a)\hspace{120pt}(b)}\\		 
   	\caption{Network usage (a) With and (b) Without Blockchain} \label{fig:netuse}
	\end{center}
\end{figure}

Network usage in different settings of FogBus are presented in Fig. \ref{fig:netuse}. In this experiment, Fog Only setting provides improved performance than Cloud Only and Integrated Fog-Cloud case, since it solely utilizes the local networking resources. The disabled Blockchain features also reduces the network usage as less amount of security attributes are required to be transferred across the infrastructures. However, network usage gets elevated when continuously tasks are generated and their associated data and information are exchanged. In this case, tuning of subsequent task generation interval can reduce the network usage to a certain scale. Thus, these adjustments make FogBus operational even when less amount of network resources are allocated for a particular IoT-enabled system.     
\subsubsection{Energy}
\begin{figure}[h]
	\begin{center}
	\centering
		\includegraphics[width=35mm, height=38mm]{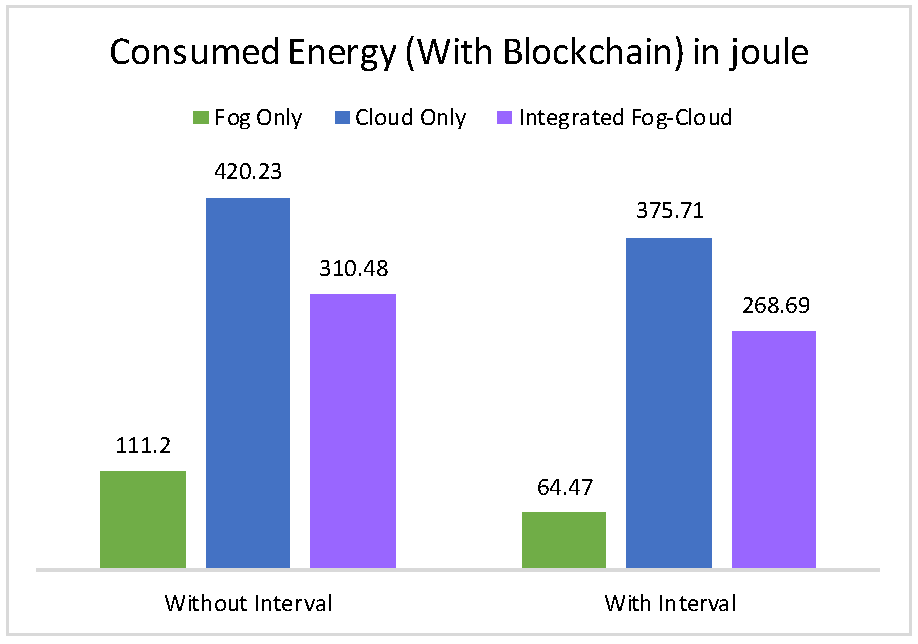}
		\hspace{20pt}
		\includegraphics[width=35mm, height=38mm]{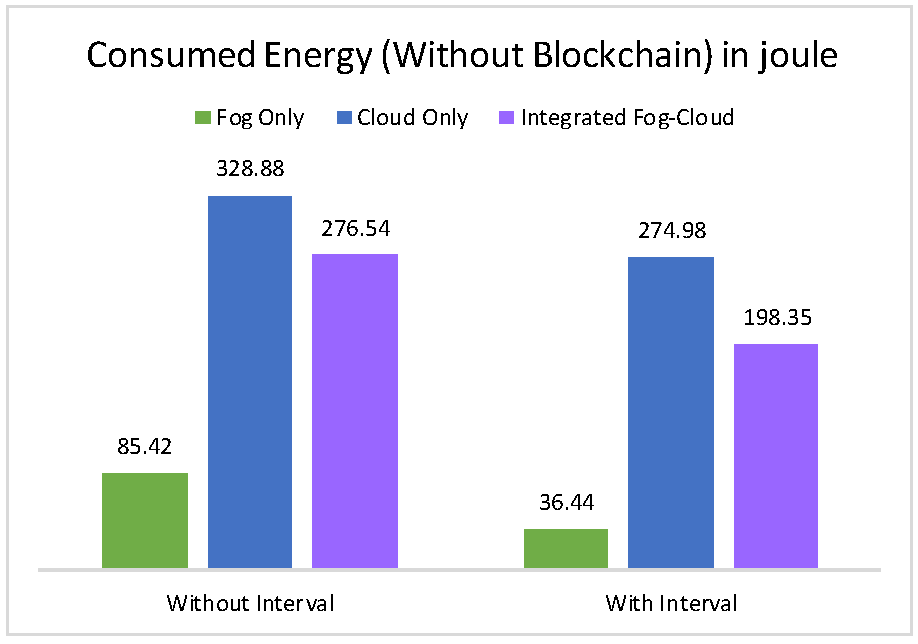}\\
        \footnotesize{(a)\hspace{120pt}(b)}\\		 
   	\caption{Energy consumption (a) With and (b) Without Blockchain} \label{fig:energy}
	\end{center}
\end{figure}

Fig. \ref{fig:energy} presents how different settings of FogBus influence energy consumption of the infrastructure. In Cloud Only setting the Fog nodes are used for networking and Cloud VMs conduct the computation whereas in Fog Only setting both the networking and computation are handled by Fog nodes. Nevertheless in Integrated Fog-Cloud case computational tasks are distributed to both the infrastructures according to the context of the system. Since, Cloud VMs consume much more energy compared to the Fog nodes, in Fog Only setting less energy is required to conduct the operations. Besides, to manage the Blockchain feature of the FogBus, additional energy is devoured. In this case, disabled Blockchain feature saves some energy for FogBus. In addition, energy consumption of an infrastructure during busy time is higher compared to its idle time. Therefore, interval between subsequent task creation assists to improve the energy usage of the infrastructure. However, it leads FogBus to process less number of tasks which can be overcome by efficient tuning of the interval time. However, these configurations help FogBus to execute applications under the energy constraints.

\begin{figure}[!t]
	\begin{center}
		\includegraphics[width=28mm, height=38mm]{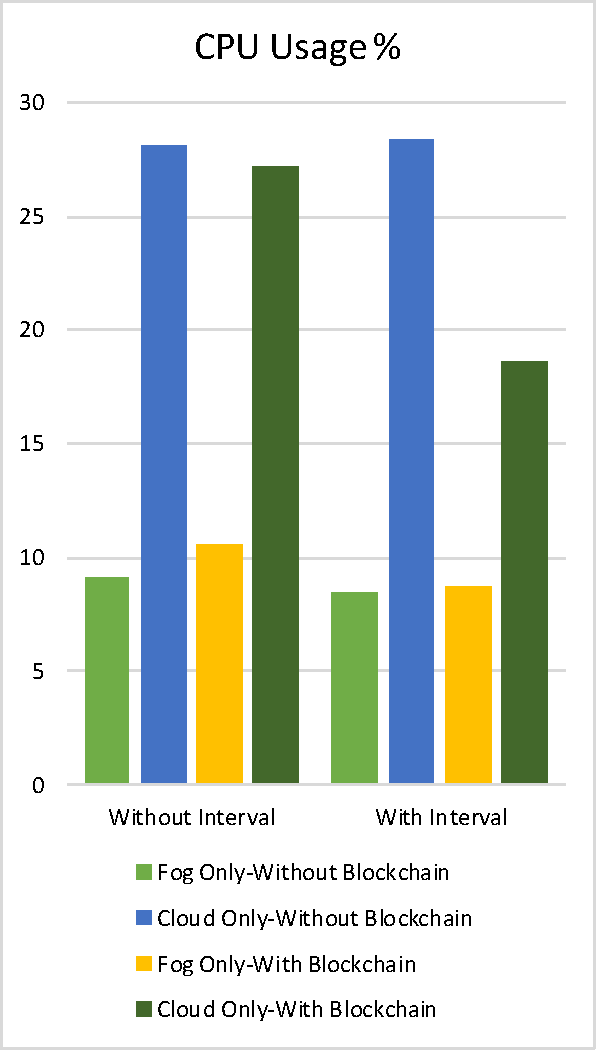}
		\hspace{1pt}
		\includegraphics[width=28mm, height=38mm]{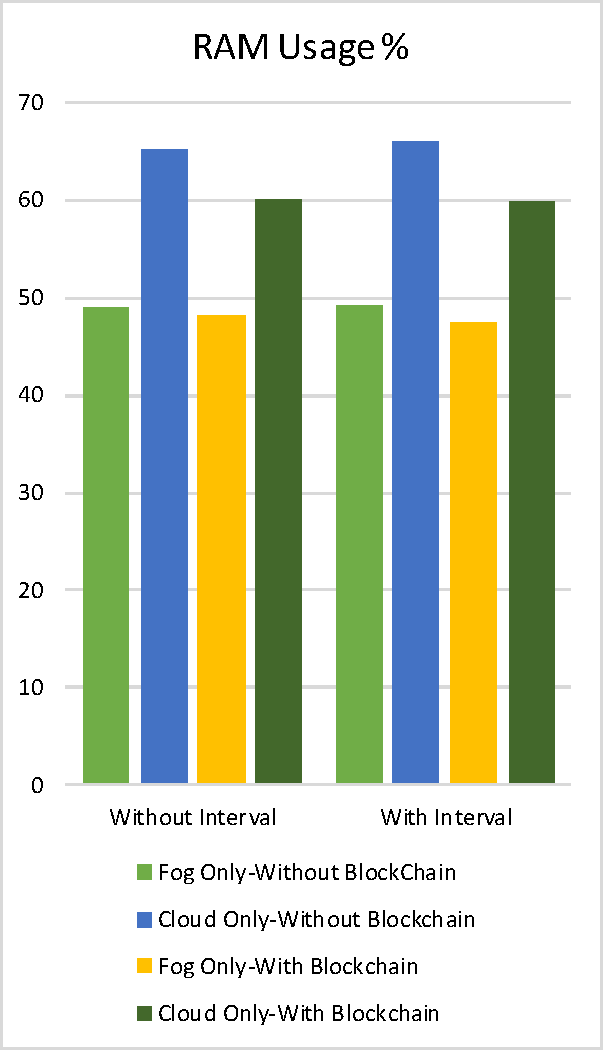}
		\hspace{1pt}
		\includegraphics[width=28mm, height=38mm]{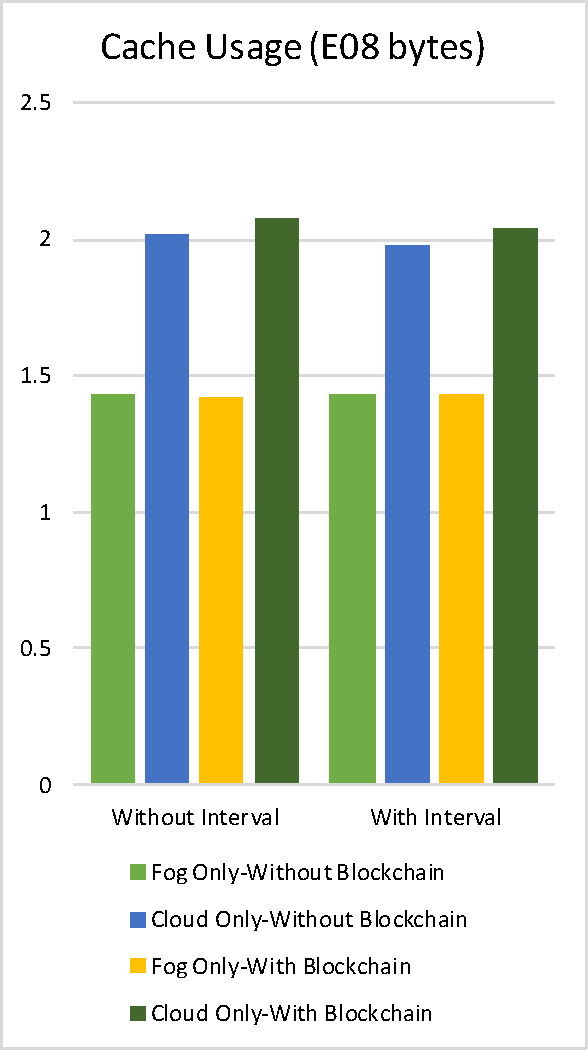}\\
        \footnotesize{(a)\hspace{60pt}(b) \hspace{60pt}(c)}\\		 
   	\caption{(a) CPU, (b) RAM and (c) Cache utilization of master} \label{fig:crc}
	\end{center}
\end{figure}
\subsubsection{CPU, RAM, Cache Usage of Broker / Master}
The Fig. \ref{fig:crc} shows the CPU, RAM and Cache usage of broker/master for various FogBus settings such as Fog Only-Without Blockchain, Cloud Only-Without Blockchain, Fog Only-With Blockchain and Cloud Only-With Blockchain. The parameter values for Fog Only setting are much lower compared to Cloud Only case since it reduces the overhead of running Cloud Plugins and storing the Cloud communication attributes. Even Without Blockchain, these parameter values also decrease as hash and proof-of-work creation for each data block are eliminated. Moreover, on any resource constrained master, these settings can ensure acceptable performance of the FogBus framework. However, since the software components of FogBus do not release their allocated resources after operations by themselves, the RAM and the cache usage for master in all settings remain almost same. In this case, the interval in subsequent task generation does not improve the state of resources. It is left to be resolved in the next version of FogBus.        

\section{Conclusion and Future Works} \label{future}
In this work, we propose the FogBus framework that can integrate different IoT-enabled systems to both Fog and Cloud infrastructures. The framework is lightweight and can harness both edge and remote resources for IoT application deployment, monitoring and management. FogBus is developed in cross platform programming languages that helps to overcome the heterogeneity of the infrastructure during application execution and end-to-end interaction. Additionally, the FogBus framework functions as a Platform-as-a-Service (PaaS) model for integrated Fog Cloud environment that not only assists application developers to build different types of IoT applications but also supports users to customize the services, and service providers to manage the resources according to the context of the system. Since some IoT-enabled systems such as health monitoring and utility service metering deal with sensitive data, FogBus applies authentication for data privacy and Blockchain for data integrity. To procure data transfer across less secure network, encryption techniques are applied in FogBus. Based on the principles of FogBus, a cost efficient prototype for Sleep Apnea analysis is also developed in this work. Applying different FogBus settings on the prototype, it is demonstrated that FogBus performs well even when large number of tasks are required to be processed, the execution of tasks are latency sensitive, network resources are not abundant, energy usage is restricted and computing instances are not resource enriched. 
\par Although FogBus is capable of enhancing service quality across diverse infrastructures,  it can be still improved in a larger scope under the following aspects.
\par \textbf{\textit{Resource management policies:}} FogBus provides flexibility to apply customized provisioning polices while allocating resources for different applications. Dynamic resource management policies on top of existing static management policy can be developed targeting load balancing among the computing infrastructures and the QoS enhancement.  
\par \textbf{\textit{Fog infrastructure virtualization:}} FogBus assists integration of Fog and Cloud computing with IoT-enabled systems. Although Cloud computing can be virtualized, in depth exploration is required to virtualize the Fog infrastructure in FogBus.  
\par \textbf{\textit{Artificial Intelligence:}} Currently FogBus does not support any artificial intelligence techniques for controlling the operations in different infrastructure and improving the resilience of the system. Inclusion of Artificial Intelligence techniques can be a significant contribution towards FogBus. 
\par \textbf{\textit{Application placement techniques:}} FogBus inherently supports distributed application execution. While placing applications in distributed manner, service latency, user expectations and deployment cost become predominant. In this case, different efficient application placement techniques can be added to the software stack of FogBus. 
\par \textbf{\textit{Runtime application migration:}} Migration of applications during runtime is very crucial if any anomaly is predicted. Different runtime application migration strategies for FogBus can be developed to handle such uncertain events.         
\par \textbf{\textit{Lightweight security features:}} Existing security features of FogBus require comparatively higher computational assistance. This consequently affects the service delivery latency, energy and network usage. Therefore, lightweight but effective security features can be helpful for further uplift of FogBus.\\    
\\The FogBus software along with source codes, users and developers manual is available from \url{https://github.com/Cloudslab/FogBus}

\bibliographystyle{elsarticle-num}

\bibliography{FogBusBib}

\end{document}